\documentclass[letterpaper, 10 pt, conference]{ieeeconf}  % Comment this line out if you need a4paper
\usepackage{graphicx}
\usepackage{amsfonts}
\usepackage{amsmath}
\usepackage{url}
\usepackage{cite}

\newcommand{\bs}{\mathbf{s}}
\newcommand{\ba}{\mathbf{a}}
\newcommand{\bx}{\mathbf{x}}
\newcommand{\bA}{\mathbf{A}}
\newcommand{\bD}{\mathbf{D}}
\newcommand{\bW}{\mathbf{W}}
\newcommand{\bI}{\mathbf{I}}
\newcommand{\reb}{\mbox{\scriptsize reb}}

\IEEEoverridecommandlockouts                              % This command is only needed if 
\title{\LARGE \bf
Graph Neural Network Reinforcement Learning for\\ Autonomous Mobility-on-Demand Systems
}

\author{Daniele~Gammelli$^{1}$, Kaidi~Yang$^{2}$, James~Harrison$^{2}$, Filipe~Rodrigues$^{1}$, Francisco~C.~Pereira$^{1}$, Marco~Pavone$^{2}$% 
\thanks{The authors would like to thank M. Zallio for help with the graphics. This research was partially supported by the Toyota Research Institute (TRI). K. Yang would like to acknowledge the support of the Swiss  National  Science  Foundation (SNSF) Postdoc.Mobility Fellowship (P400P2\_199332). This article
solely reflects the opinions and conclusions of its authors and not TRI, SNSF, or any other entity.}% 
\thanks{$^{1}$Technical University of Denmark, DK
         {\tt\small \{daga, rodr, camara\}@dtu.dk}}%
 \thanks{$^{2}$Stanford University, USA
         {\tt\small \{kaidi.yang, jharrison, pavone\}@stanford.edu}}%
}

\allowdisplaybreaks

\begin{document}

\maketitle
\thispagestyle{empty}
\pagestyle{empty}

%%%%%%%%%%%%%%%%%%%%%%%%%%%%%%%%%%%%%%%%%%%%%%%%%%%%%%%%%%%%%%%%%%%%%%%%%%%%%%%%
\begin{abstract}
Autonomous mobility-on-demand (AMoD) systems represent a rapidly developing mode of transportation wherein travel requests are dynamically handled by a coordinated fleet of robotic, self-driving vehicles.
Given a graph representation of the transportation network - one where, for example, nodes represent areas of the city, and edges the connectivity between them - we argue that the AMoD control problem is naturally cast as a node-wise decision-making problem.
In this paper, we propose a deep reinforcement learning framework to control the rebalancing of AMoD systems through graph neural networks.
Crucially, we demonstrate that graph neural networks enable reinforcement learning agents to recover behavior policies that are significantly more transferable, generalizable, and scalable than policies learned through other approaches.
Empirically, we show how the learned policies exhibit promising zero-shot transfer capabilities when faced with critical portability tasks such as inter-city generalization, service area expansion, and adaptation to potentially complex urban topologies.
\end{abstract}

%%%%%%%%%%%%%%%%%%%%%%%%%%%%%%%%%%%%%%%%%%%%%%%%%%%%%%%%%%%%%%%%%%%%%%%%%%%%%%%%

\section{INTRODUCTION}
\label{sec:introduction}
Personal urban mobility is currently dominated by the increase of private cars for \emph{fast} and \emph{anytime} point-to-point travel within cities.
However, this paradigm is currently challenged by a variety of impellent factors, such as the production of greenhouse gases, dependency on oil, and traffic congestion, especially in densely populated areas.
With the urban population projected to reach 60 percent of the world population by 2030 \cite{UN2014}, private cars are widely recognized as unsustainable for the future of personal urban mobility.
In light of this, cities face the challenge of devising services and infrastructure that can sustainably match the growing mobility needs and reduce environmental harm.

In order to address this problem, any potential solution will likely need to work towards the convergence of a variety of emerging technologies \cite{Pavone2015}.
To this regard, one of the most promising strategies is the concept of \emph{mobility-on-demand} (MoD), in which customers typically request a one-way ride from their origin to a destination and are served by a shared vehicle belonging to a larger fleet.
One of the major limitations of the MoD paradigm lies in the spatio-temporal nature of urban mobility, such that trip origins and destinations are asymmetrically distributed (e.g., commuting into a downtown in the morning and vice-versa in the evening), making the overall system \emph{imbalanced} and sensitive to disturbances.
Related to this problem, the advancement in autonomous driving technologies offers a potential solution. 
Specifically, autonomous driving could enable an MoD operator to coordinate vehicles in an automated and centralized manner, thus eliminating the need for manual intervention from a human driver.
However, controlling AMoD systems potentially entails the routing of thousands of robotic vehicles within complex transportation networks, thus effectively making the AMoD control problem an open challenge.

\begin{figure}[t]
      \centering
      \includegraphics[clip, trim=0.45cm 0.45cm 0cm 0cm,width=0.49\textwidth]{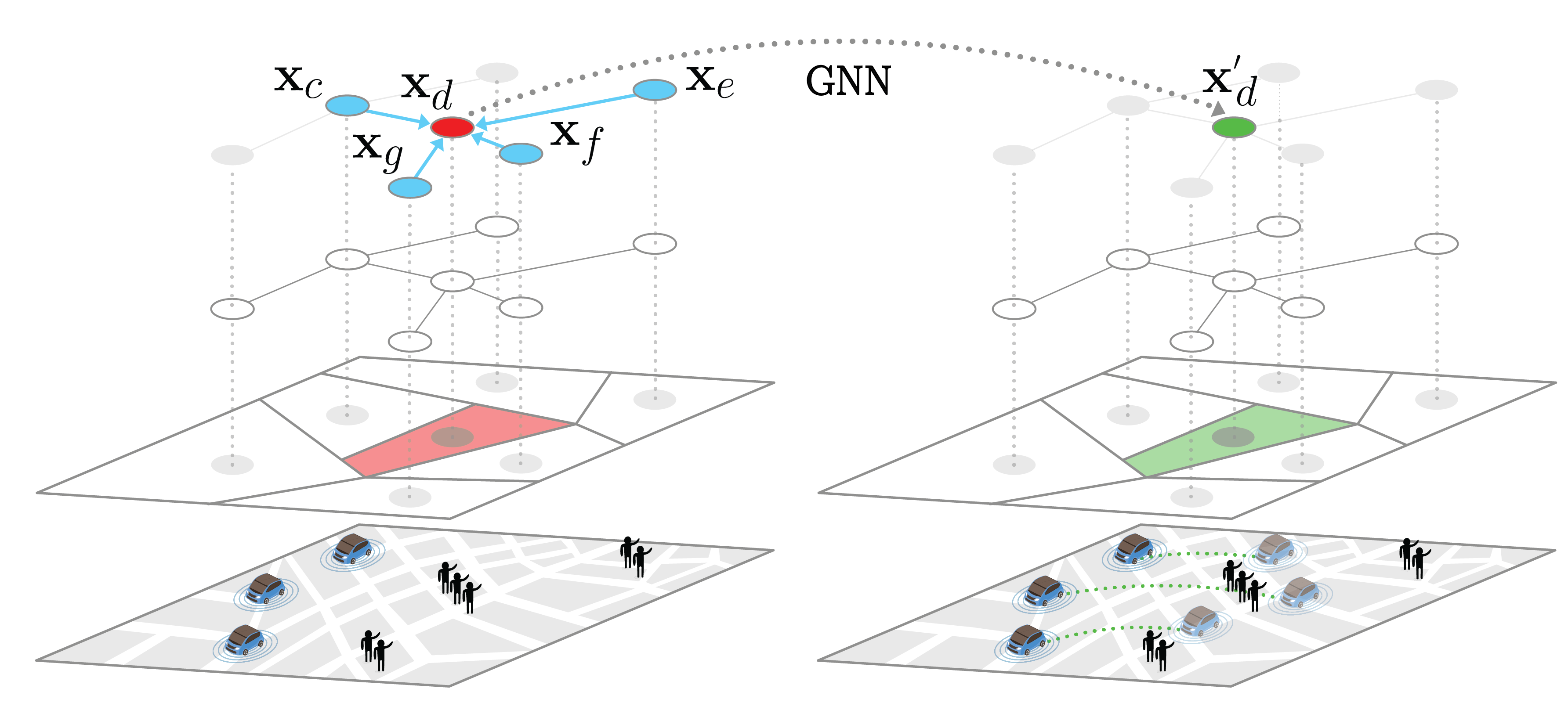}
      \caption{This paper proposes a framework to control AMoD systems by learning a shared rebalancing rule across all areas (nodes) in the transportation network. Through the use of graph convolutions, the proposed architecture aggregates information coming from both local information ($\mathbf{x}_d$, in red), as well as information about neighboring areas (in blue), to learn an updated representation ($\mathbf{x}'_d$, in green) for downstream rebalancing tasks.}
      \label{fig:gnn-for-amod}
   \end{figure}
   
In this work, we propose the use of graph neural networks to centrally control AMoD systems.
In particular, given a graph representation of the transportation network - a graph where nodes represent areas of the city and edges the connectivity between them \cite{ZhangPavone2016} (Fig.\ref{fig:gnn-for-amod}) - we learn a node-wise rebalancing policy through deep reinforcement learning. 
We argue that graph neural networks exhibit a number of desirable properties, and propose an actor-critic formulation as a general approach to learn proactive, scalable, and transferable rebalancing policies.  

%\smallskip
\noindent \textbf{Related work.} Existing literature on the real-time coordination of AMoD systems can be classified into three categories. 
The first category applies simple rule-based heuristics \cite{HylandMahmassani2018,LevinKockelmanEtAl2017} that, although efficient, can rarely yield close-to-optimal solutions. 
The second category designs Model Predictive Control (MPC) approaches based on network flow models~\cite{ZhangPavone2016,IglesiasRossiEtAl2018}, whereby an embedded open-loop optimization problem is solved at each time step to yield a sequence of control actions over a receding horizon, but only the first control action is executed. These embedded optimization problems are typically formulated into large-scale linear or integer programming problems, which may not scale well for complex AMoD networks. 

The third and most relevant category for this work employs learning-based approaches to devise efficient algorithms without significantly compromising optimality \cite{LeiQianEtAl2020,GueriauCugurulloEtAl2020,HollerVuorioEtAl2019,FluriRuchEtAl2019}.
Gueriau et al. \cite{GueriauCugurulloEtAl2020} developed RL-based decentralized approaches where the action of each vehicle is determined independently through a Q-learning policy. Although these approaches are computationally efficient, the system performance may be sacrificed due to the lack of coordination.  
Holler et al. \cite{HollerVuorioEtAl2019} developed a cooperative multi-agent approach for order dispatching and vehicle rebalancing using Deep Q-Networks \cite{MnihKavukcuogluEtAl2015} and Proximal Policy Optimization \cite{SchulmanWolskiEtAl2017}. Nevertheless, the action of only one vehicle is determined at each time step due to computation tractability, which might lead to myopic solutions. 
Fluri et al. \cite{FluriRuchEtAl2019} developed a cascaded Q-learning approach to operate AMoD systems in a centralized manner. The cascaded structure significantly reduces the number of state-action pairs, allowing more efficient learning. However, by only considering the current vehicle distribution, this approach may not perform well in scenarios with time-varying travel demand and dynamic traffic conditions, where taking reactive decisions might be sub-optimal. 
Overall, although the aforementioned RL-based works cover a wide range of algorithms, there lacks a discussion on how to define neural network architectures able to exploit the graph structure present in urban transportation networks.
In this work, we argue that exploiting the graph structure can represent a new direction for addressing relevant challenges such as transferring learning beyond the training conditions and learning from small amounts of experience.

\smallskip
\noindent \textbf{Paper contributions.} The contributions of this paper are threefold.
First, we define a novel neural architecture for a deep reinforcement learning agent, whose policy leverages the relational representative power of graph neural networks. 
By doing so, the agent can exploit the connectivity encoded by the transportation network, which is advantageous in learning a policy that is more effective, scalable, and generalizable.
Specifically, graph neural networks enable RL agents to propagate information between different regions of the transportation network before iteratively computing a rebalancing decision for all areas.

Second, we show how the proposed framework achieves close-to-optimal performance when evaluated against both control-based and learning-based approaches on real trip data from the cities of Chengdu, China and New York, USA.

Third, this work highlights how graph neural networks exhibit a series of desirable properties of fundamental practical importance for any system operator.
In particular, to the best of the authors' knowledge, this work contains the first rigorous analysis of the transfer and generalization capabilities of deep reinforcement learning approaches to the AMoD rebalancing problem. 
Specifically, results show interesting zero-shot transferability of a trained agent in the context of (i) inter-city generalization (i.e., the agent is trained on one city and directly applied to another), (ii) service area expansion (i.e., the agent is trained on a specific sub-graph and directly applied to additional areas of the city), and (iii) adaptation to irregular geographies (i.e., we measure the performance of the proposed framework when applied to arbitrarily complex transportation networks).

\section{BACKGROUND}
\label{sec:background}
In this section, we introduce the notation and theoretical foundations underlying our work in the context of Reinforcement Learning (Section \ref{subsec:the_reinforcement_learning_problem}) and Graph Neural Networks (Section \ref{subsec:graph_neural_networks}).

\subsection{The Reinforcement Learning Problem}
\label{subsec:the_reinforcement_learning_problem}
Reinforcement learning addresses the problem of learning to control a dynamical system from experience.
In this work, we will refer to a dynamical system as being entirely determined by a (fully-observed) Markov decision process (MDP) $\mathcal{M} = \left(\mathcal{S}, \mathcal{A}, P, d_0, r, \gamma \right)$, where $\mathcal{S}$ is a set of states $\bs \in \mathcal{S}$, which may be either discrete or continuous, $\mathcal{A}$ is a set of possible actions $\ba \in \mathcal{A}$, also discrete or continuous, $P$ describes the dynamics of the system through a conditional probability distribution of the form $P(\bs_{t+1} | \bs_t, \ba_t)$, $d_0$ defines the initial state distribution $d_0(\bs_0)$, $r : \mathcal{S} \times \mathcal{A} \rightarrow \mathbb{R}$ defines a reward function, and $\gamma \in (0,1]$ is a scalar discount factor.
From a reinforcement learning perspective, the final goal is to learn a policy defining a distribution over possible actions given states, $\pi(\ba_t | \bs_t)$.
We will use the term trajectory to refer to a sequence of states and actions of length $H$, given by $\tau = (\bs_0, \ba_0, \ldots , \bs_H, \ba_H)$, where $H$ may be infinite.
From these definitions, we can derive the \emph{trajectory distribution} $p_{\pi}$ for a given MDP $\mathcal{M}$ and policy $\pi$ as $p_{\pi}(\tau) = d_0(\bs_0) \prod_{t=0}^H \pi(\ba_t | \bs_t) P(\bs_{t+1} | \bs_t, \ba_t)$.
The reinforcement learning objective $J(\pi)$ can then be written as an expectation under this trajectory distribution as $J(\pi) = \mathbb{E}_{\tau \sim p_{\pi}(\tau)} \left[\sum_{t=0}^{H} \gamma^t r(\bs_t, \ba_t) \right]$.
Intuitively, the objective $J(\pi)$ provides us with a mathematical formalism for learning-based control through the maximization of the expected sum of cumulative rewards under the trajectory distribution.

At a high level, most reinforcement learning algorithms follow the same basic learning loop: the agent interacts with the MDP $\mathcal{M}$ (i.e. the environment) by applying some sort of \emph{behavior policy}, which may or may not match $\pi(\ba | \bs)$, by observing some state $\bs_t$, choosing an action $\ba_t$, and then observing the next state $\bs_{t+1}$ together with a scalar reward feedback $r_t = r(\bs_t, \ba_t)$.
This procedure may repeat for multiple steps, during which the agent uses the observed transitions $(\bs_t, \ba_t, \bs_{t+1}, r_t)$ to update its policy.

One of the most direct ways to optimize the reinforcement learning objective is to estimate its gradient and cast the learning process as approximate gradient ascent on $J(\pi)$. 
In this work, we will assume that the policy is parametrized by some parameter vector $\theta$, thus given by $\pi_{\theta}(\ba_t | \bs_t)$.
From these definitions, we can express the gradient of the objective $J(\pi)$ \cite{Williams1992} with respect to $\theta$ as $\nabla_{\theta}J(\pi_{\theta}) = \mathbb{E}_{\tau \sim p_{\pi_{\theta}} (\tau)} \big[\sum_{t=0}^H \gamma^t \nabla_{\theta} \log \pi_{\theta}(\ba_t | \bs_t) \hat{A}(\bs_t, \ba_t) \big]$, where the \emph{advantage} estimator $\hat{A}(\bs_t, \ba_t) = \sum_{t' = t}^H \gamma^{t' - t} r(\bs_{t'}, \ba_{t'}) - b(\bs_t)$ can itself be parametrized by a separate neural network, often referred to as \emph{critic}, and where the baseline $b(\bs_t)$ is typically defined as an estimator for the state value function $V^{\pi}(\bs_t) = \mathbb{E}_{\tau \sim p_{\pi}(\tau | \bs_t)}\left[\sum_{t' = t}^H \gamma^{t' - t} r(\bs_{t'}, \ba_{t'})\right]$ \cite{SuttonBarto1998}.

\subsection{Graph Neural Networks}
\label{subsec:graph_neural_networks}
One of the key reasons for the success of deep neural networks is their ability to leverage statistical properties of the data such as stationarity and compositionality through learnable hierarchical processing.
In particular, Convolutional Neural Networks (CNNs) \cite{LeCunBoserEtAl1989} have been extremely successful on a variety of vision tasks.
At their core, CNNs exploit the underlying grid-like data representation (i.e. pixels in an image) by efficiently reusing a set of local learnable filters and applying them to all input positions.
Recently, there has been a rising interest in trying to extend these concepts to non-Euclidean geometric data, such as graphs \cite{BronsteinBrunaEtAl2017}.

In order to extend the concept of convolutions to graph structured data, a key property which needs to be satisfied is the one of \emph{permutation invariance}. 
For illustration, consider a graph representation of the transportation network comprised of $n$ nodes, whose attributes are denoted by $\{\bx_1, \bx_2, \ldots, \bx_n\}$. 
We will refer to a computation as permutation invariant if its output is independent of the node ordering.
On the other hand, a non-permutation invariant computation would consider each ordering as fundamentally different, and thus require an exponential number of input/output training examples before being able to generalize.

In this work, we argue that permutation-invariant computational models operating on irregular graphs---such as those enabled by graph neural networks---pair naturally with the predominant modelling techniques in transportation engineering broadly and AMoD in particular. 
At the heart of our claim is the observation that areas in a city, just like nodes in a graph, do not have a natural order.
Instead, graphical representations of transportation systems are naturally defined by node and edge properties such as demand in a region or travel time between regions.
An effective control policy should not be affected by the order in which we consider the areas, but rather solely by the properties (i.e. attributes) of those areas.

Given a graph $\mathcal{G} = (\mathcal{V}, \mathcal{E})$, where $\mathcal{V} = \{v_i\}_{i=1:N_v}$ and $\mathcal{E} = \{e_k\}_{k=1:N_e}$ respectively define the sets of nodes and edges of $\mathcal{G}$, most current graph neural network models can be seen as methods attempting to learn a permutation-invariant function taking as input (i) a $D$-dimensional feature description $\mathbf{x}_i$ for every node $i$ (typically summarized in a $N_v \times D$ feature matrix $\mathbf{X}$), (ii) a representative description of the graph structure in matrix form $\bA$ (typically in the form of an adjacency matrix), and produce an updated representation $\mathbf{x'}_i$ for all nodes in the graph.

An architecture of particular interest for this work is the Graph Convolution Network (GCN) \cite{KipfWelling2017}.
At its core, a graph convolutional operator describes a parametric function $f(\mathbf{X}, \bA)$ for efficient information propagation on graphs.
Specifically, a GCN defines the following propagation rule:
\begin{equation}
    \mathbf{X'} = f(\mathbf{X}, \bA) = \sigma\left(\hat{\bD}^{-\frac{1}{2}} \hat{\bA} \hat{\bD}^{-\frac{1}{2}} \mathbf{X} \bW\right), 
    \label{eq:gcn}
\end{equation}
with $\hat{\bA} = \bA + \bI$, where $\bI$ is the identity matrix, $\hat{\bD}$ is the diagonal node degree matrix of $\hat{\bA}$, $\sigma(\cdot)$ is a non-linear activation function (e.g., ReLU) and  $\bW$ is a matrix of learnable parameters.
Intuitively, graph convolutions describe a permutation-invariant propagation rule which updates node-level features by aggregating information from neighboring nodes through the parameterization of a shared local filter across all locations in the graph (Fig. \ref{fig:gnn-for-amod}).

\begin{figure*}[t]
      \centering
     \includegraphics[width=\textwidth]{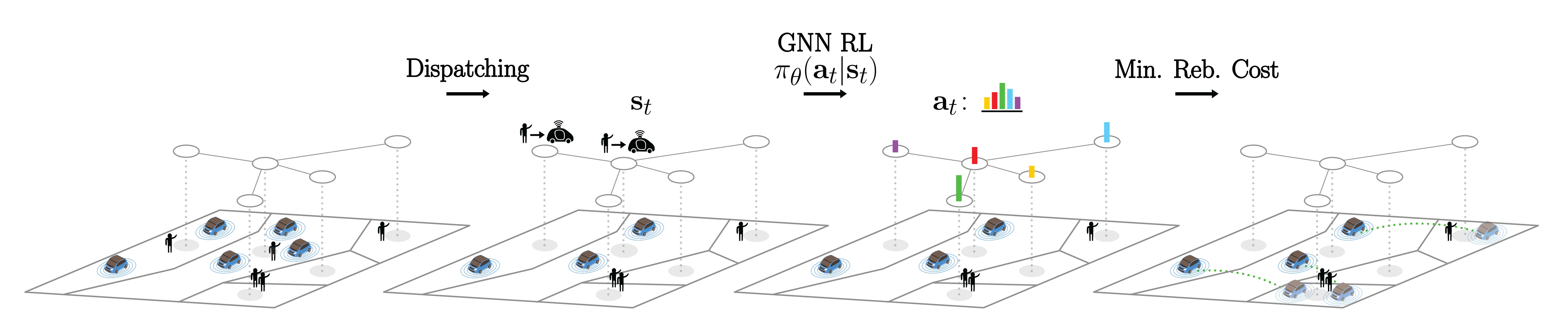}
      \caption{An illustration of the three-step framework determining the proposed AMoD control strategy.
      Given the current distribution of idle vehicles (cars) and user transportation requests (stick figures), the control strategy is defined by: (1) dispatching idle vehicles to specific trip requests by solving a matching problem (thus, characterizing the current state $\bs_t$ of the system), (2) computing an action $\ba_t$ (i.e. the desired distribution of idle vehicles) using some policy $\pi_{\theta}$, and (3) translate $\ba_t$ into actionable rebalancing trips such that overall rebalancing cost is minimized.}
      \label{fig:3-step-framework}
   \end{figure*}

\section{RELATIONAL DEEP RL FOR AMOD SYSTEMS}
\label{sec:relational_deep_rl_for_amod_systems}
In this section, we propose a control framework to learn effective AMoD rebalancing policies from experience.
Towards this aim, we consider a transportation network represented by a complete graph $\mathcal{G} = (\mathcal{V}, \mathcal{E})$ with $M$ single-occupancy vehicles, where $\mathcal{V}$ represents the set of stations (e.g., pick-up or drop-off locations) and $\mathcal{E}$ represents the shortest paths connecting the stations. Let us denote $N_v = |\mathcal{V|}$ as the number of stations. 
The time horizon is discretized into a set of discrete intervals $\mathcal{T}=\{1,2,\cdots, T\}$ of a given length $\Delta T$. 
The travel time for edge $(i,j)\in\mathcal{E}$ is defined as the number of time steps it takes a vehicle to travel along the shortest path between station $i$ and station $j$, denoted as an integer $\tau_{ij} \in \mathbb{Z}_+$.
Further denote $c_{ij}$ as the cost of traveling through an edge $(i,j) \in \mathcal{E}$, which can be calculated as a function of travel time $\tau_{ij}$. 

At each time step, customers arrive at their origin stations and wait for vehicles to transport them to their desired destinations.
The trip traveling from station $i\in\mathcal{V}$ to station $j \in \mathcal{V}$ at time step $t$ is characterized by demand $d_{ij}^t$ and price $p_{ij}^t$. Consequently, passengers departing from origin station $i$ at time $t$ will arrive at the destination station $j$ at time $t+\tau_{ij}$.  

The AMoD operator coordinates a fleet of taxi-like fully-autonomous vehicles to serve the transportation demand. The operator matches passengers to vehicles, and the matched vehicles will deliver passengers to their destinations. Let us denote $x_{ij}^t\leq d_{ij}^t$ as the passenger flow, i.e. the number of passengers traveling from station $i$ to station $j$ at time step $t$ that are successfully matched with a vehicle. Passengers not matched with any vehicles will leave the system. For vehicles not matched with any passengers, the operator will either have them stay at the same station or rebalance them to other stations. Let us denote $y_{ij}^t$ as the rebalancing flow, i.e., the number of vehicles rebalancing from station $i$ to station $j$ at time step $t$. 

\smallskip
\textbf{Remark.} The following remarks are made in order. First, we assume travel times are given and independent of the control of the AMoD fleet. This assumption applies to cities where the AMoD fleet constitutes a relatively small proportion of the entire vehicle population on the transportation network, and thus the impact of the AMoD fleet on traffic congestion is marginal. This assumption can be relaxed by training the proposed RL model in an environment considering the endogenous congestion caused by the AMoD fleet, which will be investigated as future work. Second, without loss of generality, we assume that the arrival process of passengers for each OD pair is a time-dependent Poisson process. We further assume that such process is independent of the arrival processes of other OD pairs and the coordination of AMoD fleets. These assumptions are commonly used to model transportation requests \cite{Daganzo1978}. The proposed approach, nevertheless, can be readily extended to consider other types of arrival processes.  

\subsection{A Three-Step Framework}
\label{subsec:three-step-framework}
As illustrated in Figure~\ref{fig:3-step-framework}, we adopt a three-step decision-making framework similar to \cite{FluriRuchEtAl2019} to control an AMoD fleet: (1) deriving passenger flow by solving a matching problem, (2) computing the desired distribution of idle vehicles at the current time step by using the learned policy $\pi_{\theta}(\ba_t | \bs_t)$, (3) converting the desired distribution to rebalancing flow by solving a minimal rebalancing-cost problem. Notice that in the three-step procedure, we have an action at each node as opposed to along each edge (as in the majority of literature). In other words, we reduce the dimension of the action space of the AMoD rebalancing MDP to $N_{v}$ (compared to $N_{v}^2$ in edge-based approaches), and thus significantly improve scalability of training and implementation.

We now explain in more detail the three-step decision framework that the AMoD operator employs at each time step $t$. The first step is passenger matching, wherein the following matching problem is solved to derive passenger flows $\{x_{ij}^t\}_{i,j\in\mathcal{V}}$: 
\begin{subequations}
\begin{align}
    \max_{\{x_{ij}^t\}_{i,j\in\mathcal{V}}} \quad& \sum_{i,j\in\mathcal{V}}x_{ij}^t(p_{ij}^t- c_{ij}^t) \label{eq:matching_obj}\\
    \rm{s.t.} \,\,\,\,\, \quad& 0\leq x_{ij}^t \leq d_{ij}^t, ~i,j\in\mathcal{V}, \label{eq:matching_con}  
\end{align}\label{eq:matching}
\end{subequations}
where the objective function Eq.(\ref{eq:matching_obj}) represents the total profit of passenger assignment calculated as the difference between revenue and cost, and the constraint Eq.(\ref{eq:matching_con}) ensures that the passenger flow is non-negative and does not exceed the demand. Notice that since the constraint matrix is totally unimodular, the resulting passenger flows are positive integers, i.e., $x_{ij}^t\in \mathbb{Z}_+$ if $d_{ij}^t \in \mathbb{Z}_+,~\forall i,j\in\mathcal{V}$. 

The second step entails determining the desired idle vehicle distribution  $ \ba_{\reb}^t = \{\ba_{\reb, i}^t\}_{i \in \mathcal{V}}$, where $\ba_{\reb, i}^t \in [0,1]$ defines the percentage of currently idle vehicles to be rebalanced towards station $i$ in time step $t$, and $\sum_{i \in \mathcal{V}} \ba_{\reb, i}^t = 1$. 
With desired distribution $\ba_{\reb}^t$, denote $\hat{m}_i^t=\lfloor a_{reb,i}^t\sum_{i\in\mathcal{V}}{m_i^t}\rfloor$ as the number of desired vehicles, where $m_i^t$ represents the actual number of idle vehicles in region $i$ at time step $t$. Here, the floor function $\lfloor\cdot\rfloor$ is used to ensure that the desired number of vehicles is integer and always available ($\sum_{i\in\mathcal{V}}\hat{m}_i^t \leq \sum_{i\in\mathcal{V}}m_i^t$).

The third step entails rebalancing, wherein a minimal rebalancing-cost problem is solved to derive rebalancing flows $\{y_{ij}^t\}_{(i,j)\in \mathcal{E}}$:
\begin{subequations}
\begin{align}
    \min_{\{y_{ij}^t\}_{(i,j)\in \mathcal{E}}\in \mathbb{Z}_+^{|\mathcal{E}|}} \quad& \sum_{(i,j)\in\mathcal{E}}c_{ij}y_{ij}^t \label{eq:reb_obj}\\
    \rm{s.t.}  \quad \quad \quad \,
    & \sum_{j\neq i}(y_{ji}^t - y_{ij}^t) + m_i^t \geq \hat{m}_i^t , ~i\in\mathcal{V},\label{eq:reb_con1} \\
    \quad& \sum_{j\neq i} y_{ij}^t \leq m_i^t,~i\in\mathcal{V},\label{eq:reb_con2}
\end{align}\label{eq:reb}
\end{subequations}
where the objective function Eq.(\ref{eq:reb_obj}) represents the rebalancing cost, constraint Eq.(\ref{eq:reb_con1})  ensures that the resulting number of vehicles (the left hand size) is close to the desired number of vehicles (the right hand side), and Eq. (\ref{eq:reb_con2}) limits the rebalancing flows from a region to the vehicles available in that region. 

\subsection{The AMoD Rebalancing Markov Decision Process}
\label{subsec:amod_mdp}
In this section, we formulate the AMoD rebalancing problem as an MDP. Specifically, we attempt to learn a behavior policy to select the desired distribution of idle vehicles, as introduced in Section \ref{subsec:three-step-framework}, by defining the following MDP:
\begin{equation}
    \mathcal{M}_{\reb} = (\mathcal{S}_{\reb}, \mathcal{A}_{\reb}, P_{\reb}, r_{\reb}, \gamma).
\label{eq:reb_mdp}
\end{equation}
In what follows, we define each of the elements describing the MDP for the AMoD rebalancing problem.

\smallskip\emph{Action Space ($\mathcal{A}_{\reb}$):} Given a number of available vehicles $M_a \leq M$, and their current spatial distribution between the $N_v$ stations, we consider the problem of determining the \emph{desired idle vehicle distribution} $\ba_{\reb}^t$.
Specifically, the proposed behavior policy will have the task of describing a probability distribution over stations, indicating the percentage of idle vehicles to be rebalanced in each station.

\smallskip\emph{Reward ($r_{\reb}$):} We define the reward function in the MDP from the perspective of an AMoD operator.
That is, we express our objective so to recover behavior policies able to maximize travel demand satisfaction and provider profit, while minimizing the cost of unnecessary vehicle trips in the system. 
Specifically, each trip between two stations $i,j$ will be characterized by a price $p_{ij}$ and a cost $c_{ij}$, with $p_{ij} = 0$ in case of rebalancing trips. 
We can naturally express this objective through the following reward function:

\begin{equation}
    r_{\reb} = \sum_{i,j \in \mathcal{V}} x_{ij}^t (p_{ij}^t - c_{ij}^t) -  \sum_{(i,j) \in \mathcal{E}} y_{ij}^tc_{ij}^t,
\end{equation}
where, as introduced in Section \ref{sec:relational_deep_rl_for_amod_systems}, we denote the passenger and rebalancing flows as $x_{ij}^t$ and $y_{ij}^t$, respectively.

\smallskip\emph{State Space ($\mathcal{S}_{\reb}$):} We define the state in the rebalancing MDP to contain the information needed to determine proactive rebalancing strategies.
Specifically, this will require knowledge of the structure of the transportation network through its adjacency matrix $\bA$, together with additional station-level information by means of a feature matrix $\mathbf{X}$.
In our experiments, we choose the adjacency matrix $\bA$ to describe a transportation network where each station is connected to its spatially adjacent regions, such that all neighboring areas are connected by an edge.
Moreover, we choose the feature matrix $\mathbf{X}$ to be a collection of three main sources of information.
Firstly, we characterize the MoD system by the current availability of idle vehicles in each station $m_i^t \in [0, M], \forall i \in \mathcal{V}$.
Given a planning horizon $T$, we also consider the \emph{projected} availability of idle vehicles $\{m_i^{t^{'}}\}_{t^{'}=t, \ldots, t+T}$, where this is estimated based on previously assigned passenger and rebalancing trips.
Secondly, an effective rebalancing strategy will also depend on both current $d_{ij}^t$ and estimated $\{\hat{d}_{ij}^{t^{'}}\}_{t^{'}=t, \ldots, t+T}$ transportation demand between all stations.
In this work, we assume to have access to a noisy and unbiased estimate of demand in the form of the rate of the underlying time-dependent Poisson process describing travel behavior in the system, although this could come from a predictive model such as \cite{GammelliEtAl2020}.
Lastly, we also include provider-level information such as trip price $p_{ij}^t$ and cost $c_{ij}^t$.
By means of this definition of the state space, we provide the behavior policy with meaningful information for it to capture statistics of the current and estimated future state of the MoD system, together with operational provider information and performance.

\smallskip\emph{Dynamics ($P_{\reb}$):} The dynamics in the rebalancing MDP describe both the stochastic evolution of travel demand patterns, as well as how rebalancing decisions influence future state elements, such as the availability and distribution of idle vehicles.
Specifically, the evolution of travel demand between stations $d_{ij}^t$ is independent of the rebalancing action, and follows a time-dependent Poisson process (in our experiments, estimated from real trip travel data).
On the other hand, some of the state variable's transitions deterministically depend on the chosen action.
For example, the estimated availability $\{m_i^{t^{'}}\}_{t^{'}=t, \ldots, t+T}$ is uniquely defined as the sum of the current availability $m_i^t$ together with the projected number of incoming vehicles at time $t'$ (from both passenger and rebalancing flows), minus the vehicles currently chosen to be rebalanced.
Finally, state variables related to provider information, such as trip price $p_{ij}^t$ and cost $c_{ij}^t$ are assumed to be externally decided and known beforehand (hence, independent from the actions selected by the behavior policy). 

\subsection{Graph Neural Network RL for Rebalancing}
\label{subsec:gnn_rl}
\begin{figure}[t]
      \centering
      \includegraphics[width=0.85\linewidth]{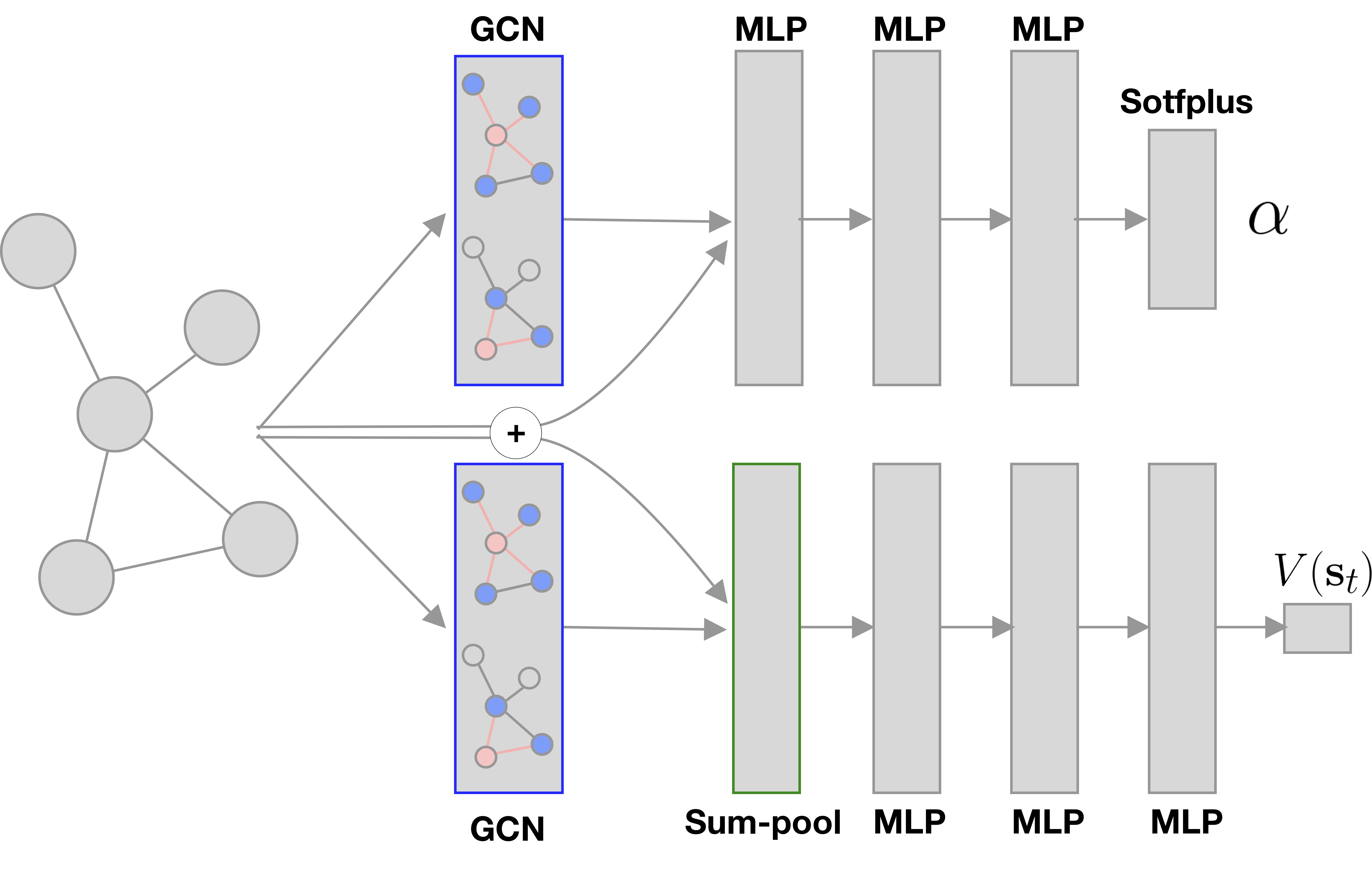}
      \caption{An illustration of the neural architecture describing the policy (top) and value function estimator (bottom).
      Intuitively, node features are first updated through the use of graph convolutions, and then passed through a series of non-linearities to compute (i) concentration parameters $\mathbf{\alpha}$ of the Dirichlet policy and (ii) an estimate of the value function $V(\bs_t).$}
      \label{fig:diagram}
  \end{figure}

Having formally defined the AMoD rebalancing problem as an MDP, this section introduces the neural architecture for the policy $\pi_{\theta}(\ba_t | \bs_t)$ and the value function estimator $V_{\phi}(\bs_t)$ characterizing the proposed Advantage Actor-Critic (A2C) algorithm \cite{MnihPuigdomenechEtAl2016}.
Fig. \ref{fig:diagram} provides a schematic illustration.

\smallskip
\noindent\textbf{Policy.} As introduced in Section \ref{subsec:amod_mdp}, a rebalancing action is defined as the desired distribution of idle vehicles across all $N_v$ stations.
    Thus, in order for $\pi_{\theta}(\ba_t | \bs_t)$ to define a valid probability density over actions, we devise the output of our policy network to represent the concentration parameters $\mathbf{\alpha} \in \mathbb{R}_{+}^{N_v}$ of a Dirichlet distribution, such that $\ba_t \sim \text{Dir}(\ba_t | \mathbf{\alpha}) = \pi_{\theta}(\ba_t | \bs_t)$ where $\text{Dir}(\cdot)$ denotes the Dirichlet distribution, and where the positivity of $\mathbf{\alpha}$ is ensured by e.g., a Softplus nonlinearity.
Concretely, the neural network used in our implementation consists of one layer of graph convolution with skip-connections and ReLU activations, whose output is then aggregated across neighboring nodes using a permutation-invariant sum-pooling function, and finally passed to three MLP layers of 32 hidden units to produce the Dirichlet concentration parameters.

\smallskip
\noindent\textbf{Value Function.}
The architecture used to define the value function $V_{\phi}(\bs_t)$ is in many ways identical to the architecture used to characterize the policy. 
The main difference between the two architectures lies in an additional \emph{global} sum-pooling performed on the output of the graph convolution.
In this way, the value function is able to aggregate information across all nodes in the graph, thus computing a single value function estimate for the entire network. 

\section{EXPERIMENTS}
\label{sec:experiments}
In this section, we present simulation results that demonstrate the performance of our proposed approach\footnote{Code available at: \url{https://github.com/DanieleGammelli/gnn-rl-for-amod}}.
Specifically, the goal of our experiments is to answer the following questions: (1) Can the proposed A2C-GNN learn effective rebalancing strategies on real-world urban mobility scenarios? (2) What are the generalization capabilities of a behavior policy learned through our approach? (3) Computationally, what are the advantages of GNN-based RL approaches compared to traditional control-based strategies?

\smallskip When possible, we compare our results to an oracle-MPC that receives perfect information of the system dynamics and future states of mobility requests.
Specifically, we compare our A2C-GNN agent with six rebalancing algorithms:

\noindent1) \emph{Equally distributed policy (ED):} at each decision, we take rebalancing actions so to recover an equal distribution of idle vehicles across all areas in the transportation network.

\noindent2) \emph{Cascaded Q-Learning (CQL) \cite{FluriRuchEtAl2019}:} rebalancing policies are learned through tabular Q learning where the current distribution of idle vehicles is chosen as the sole representation of the state space.
To achieve a more scalable state-action size, this approach introduces a hierarchical representation of the transportation network, thus learning behavior policies from coarse to finer levels.

\noindent3) \emph{A2C-MLP:} we replace the graph neural network used in our approach with a  (feed-forward) neural network consisting of four layers of $128, 64, 32,$ and $32$ hidden units, respectively. 
For this model to correctly process the data, we aggregate all node features in a single vectorial representation.

\noindent4) \emph{A2C-CNN:} we replace the graph neural network used in our approach with a  convolutional neural network consisting of two layers of $3 \times 3$ convolutions with $33$ and $44$ channels, respectively. 
In order to mantain the spatial granularity unchanged throughout the network, in our implementation we use dimensionality-preserving zero-padding.
We further process the output of the convolutional layers with two MLP layers both characterized by $128$ hidden units.
For this model to correctly process the data, we organize all node features as a spatial grid described by $D$ channels.

\noindent5) \emph{MPC-tri-level (Oracle):} we optimize the desired vehicle distribution using a Model Predictive Control (MPC) approach with a tri-level embedded optimization model with perfect foresight information, where the upper-level model determines the desired vehicle distribution, and the lower-level models are the matching problem (Eq.~\ref{eq:matching}) and the minimum rebalancing-cost problem (Eq.~\ref{eq:reb}). I.e., the MPC-tri-level approach optimizes the desired vehicle distribution considering its impact on the other modules within the three-step framework. Therefore, this approach serves as an \emph{oracle} that provides a performance upper bound for any algorithm within the three-step framework. Notice that the embedded tri-level optimization model is an NP-hard problem, thus does not scale well as the number of stations increases. 

\noindent6) \emph{MPC-standard:} we directly optimize the passenger flow and rebalancing flow using a standard formulation of MPC \cite{ZhangRossiEtAl2016b}. This approach is used to evaluate the sub-optimality of the three-step framework.  Notice that although the embedded optimization of MPC-standard is a linear programming model, it may not meet the computation requirement of real-time applications (e.g., obtaining a solution within several seconds) for large scale networks. 

These rebalancing algorithms are evaluated using two case studies inspired by New York City, USA, and the city of Chengdu, China, whereby we study a hypothetical deployment of AMoD systems to serve the morning commute demand in popular areas of Manhattan (8 a.m. -- 10 a.m., with a size of 4\,km$\times$4\,km) and  Chengdu (7 a.m. -- 10 a.m., with a size of 10\,km$\times$10\,km), respectively. 
The studied areas in both case studies are divided into grid-like blocks, each of which represents a station. 
These case studies are generated using trip record datasets provided by the NYC Taxi and Limousine Commission (March 2013)  \cite{TaxiLimousineCommission2013} and the Didi Chuxing Gaia Initiative (November 2016)  \cite{Didi2016}.
The trip records are converted to demand, travel times, and trip prices between blocks. 
Here, we consider stochastic time-varying demand patterns, whereby customer arrival is assumed to be a time-dependent Poisson process, and the Poisson rates are aggregated from the trip record data every 15 minutes. 

All models were implemented using PyTorch \cite{PaszkeGrossEtAl2019} (for the RL modules) and the IBM CPLEX solver \cite{ios_ILOG:1987} (for the dispatching and minimal rebalancing cost problems).
We train each model using stochastic gradient ascent on the RL objective $J(\pi)$ using the Adam optimizer \cite{KingmaBa2015}, with a fixed learning rate of $0.003$ for both actor and critic architectures.
We train A2C for 16,000 episodes, with discount factor $0.97$ and episode length $T=60$.
In our experiments, we care about a set of Key Performance Indicators not included within the reward function.
Specifically, we also monitor (i) \emph{Served demand}: defined as the total number of trips satisfied by the AMoD control strategy, (ii) \emph{Rebalancing cost}: defined as the overall cost induced on the system by the rebalancing policy, and (iii) \emph{Percentage deviation} from MPC-tri-level performance having oracle information of future system states (\%Dev. MPC).

\subsection{Chengdu and New York Cases}

\begin{table}[t]
\caption{System Performance on Chengdu $4 \times 4$ Network}
\centering
\begin{tabular}{p{2cm} c c c}
    \hline
     & Reward & Served & Rebalancing  \\
     & (\%Dev. MPC-tri-level) & Demand & Cost (\$) \\ [0.25ex] 
    \hline
    ED & 12,538 (-19.2\%) & 41,189 & 3,397 \\ [0.2ex]
    CQL & 12,334 (-20.5\%)  & 41,273   &  4,174 \\[0.15ex]
    A2C-MLP & 12,530 (-19.2\%) & 41,076 & 3,899 \\ [0.2ex]
    A2C-CNN & 13,274 (-14.5\%) & 40,904 & 3,087 \\ [0.2ex]
    A2C-GNN (ours) & 15,167 (\textbf{-2.2\%}) & 40,578 & 1,063 \\ [0.2ex]
    MPC-tri-level & 15,516 (0\%) & 42,425 & 1,453 \\ [0.2ex]
    MPC-standard & 16,702 (7.6\%) & 44,662 & 1,162 \\
    \hline 
    A2C-GNN-0Shot\rule{0pt}{2.0ex} & 14,791 (\textbf{-4.7\%}) & 40,646 & 1,467 \\ 
    \hline
    \end{tabular}%
  \label{tab:didi_1}%
\end{table}

\begin{table}[t]
\caption{System Performance on New York $4 \times 4$ Network}
\centering
\begin{tabular}{p{2cm} c c c}
    \hline
     & Reward & Served & Rebalancing  \\
     & (\%Dev. MPC-tri-level) & Demand & Cost (\$) \\ [0.25ex] 
    \hline
    ED & 30,746 (-10.7\%) & 8,770 & 7,990 \\ [0.2ex]
    CQL & 30,496 (-11.4\%) & 8,736   & 8,284 \\[0.15ex]
    A2C-MLP & 30,664 (-10.9\%) & 8,773 & 7,920 \\ [0.2ex]
    A2C-CNN & 30,443 (-11.5\%) & 8,904 & 8,775 \\ [0.2ex]
    A2C-GNN (ours) & 33,886 (\textbf{-1.6\%}) & 8,772 & 5,038 \\ [0.2ex]
    MPC-tri-level & 34,416 (0\%) & 8,865 & 4,647 \\ [0.2ex]
    MPC-standard & 35,356 (2.7\%) & 8,968 & 4,296 \\ [0.2ex]
    \hline 
    A2C-GNN-0Shot\rule{0pt}{2.0ex} & 33,397 (\textbf{-3.0\%}) & 8,628 & 4,743 \\ 
    \hline
    \end{tabular}%
  \label{tab:nyc_1}%
\end{table}

In our first simulation experiment, we study system performance when defining a 16-dimensional (i.e. $4 \times 4$) grid over the cities of Chengdu, China and New York, USA.
Each grid cell defines an area of approximately 6\,$\rm{km}^2$ and 1\,$\rm{km}^2$ for Chengdu and New York, respectively.
Results in Tables \ref{tab:didi_1} and \ref{tab:nyc_1} show that A2C-GNN can learn rebalancing policies able to achieve close-to-optimal system performance on both tasks.
Specifically, A2C-GNN's performance is only $2.2\%$ (Chengdu) and $1.6\%$ (New York) away from the oracle performance.
Interestingly, A2C-GNN is able to exploit its learned shared local filter to achieve more than $65\%$ (Chengdu) and $36\%$ (New York) cost savings in its rebalancing trips when compared to learning-based approaches based on different neural architectures.
Moreover, by monitoring the number of customers served, we notice how A2C-GNN learns rebalancing policies able to proactively select more profitable trips. 
Specifically, in Table \ref{tab:didi_1}, results show that A2C-GNN is able to achieve a $14\%$ increase in profit (i.e., reward) compared to the second-best non-oracle approach (A2C-CNN), despite having the lowest number of customers served across all methods.

\subsection{Inter-city Portability}
\label{subsec:inter-city_portability}
To assess the transferability and generalization capabilities of A2C-GNN, we also study the extent to which policies can be trained on one city and later applied to the other \emph{without further training} (i.e., zero-shot).
Specifically, in this simulation experiment, we select the pre-trained policy on Chengdu data, and examine its zero-shot performance when deployed in New York's transportation network, and vice-versa.
Without any fine-tuning, the only possibility of an effective rebalancing policy to emerge is for the agent to have learned a high-level, abstract and generalizable understanding of the system dynamics.
Tables \ref{tab:didi_1} and \ref{tab:nyc_1} show the adaptation performance (A2C-GNN-0Shot) when deploying the New York policy on Chengdu, and vice-versa.
The results show that that rebalancing policies recovered by independently training on a single city exhibit an interesting degree of inter-city portability.
In particular, despite the substantially diverse mobility patterns, the zero-shot policy shows only a slight drop in performance when compared to its fully re-trained counterpart: $1.4\%$ for New York and $2.5\%$ for Chengdu, respectively.
We argue that a key component in making possible this kind of inter-city generalization is the relational inductive bias typical of graph neural networks. 
In other words, we believe the explicit representation of entities (such as areas in a transportation network), together with mechanisms to compute their interaction (e.g., graph convolutions) lies at the heart of learning highly generalizable control policies.
From an application perspective, policies with structural transfer capabilities enable AMoD operators to expand their service to new cities without having to train a new control agent, thus being able to re-use actionable knowledge from other service locations.

\subsection{Service Area Expansion}
\label{subsec:service_area_expansion}

\begin{table}[t]
\caption{System Performance on New York $8\times 8$ Network}
\centering
\begin{tabular}{p{2cm} c c c}
    \hline
     & Reward & Served & Rebalancing  \\
     & (\%Dev. MPC-standard) & Demand & Cost (\$) \\ [0.25ex] 
    \hline
    ED & 41,930 (-24.4\%) & 13,028 & 11,023 \\ [0.2ex]
    A2C-GNN-0Shot & 46,516 (\textbf{-15.1\%}) & 13,974 & 10,083 \\ [0.2ex]
    A2C-GNN & 47,843 (\textbf{-12.6\%}) & 14,165 & 12,165 \\ [0.2ex]
    MPC-standard & 54,737 (0\%) & 16,275 & 10,389 \\
    \hline
    \end{tabular}%
  \label{tab:nyc_expansion}%
\end{table}

To further study how well A2C-GNN can generalize to conditions unseen during training, we now consider the case of a hypothetical service area expansion.
We define the task as follows: given 64 stations in the New York scenario (organized as a $8\times 8$ grid over the city's geography), we first learn a rebalancing policy in the inner $4 \times 4$ region, which we then use, without any fine-tuning, on the full $8 \times 8$ grid.
Despite the $300\%$ increase in service area, results in Table \ref{tab:nyc_expansion} show that A2C-GNN-0Shot is only $2.5\%$ less profitable and satisfies $1.3\%$ fewer customers when compared to its fully-retrained counterpart (A2C-GNN), thus exhibiting an interesting degree of portability to scenarios unseen during training.
Given the scalability constraints of MPC-tri-level when faced with larger sized networks, in this set of experiments we evaluate the percentage divergence with respect to MPC-standard.
From a practical perspective, it is important to underline how this experimental setting might represent a set of common real-world scenarios, such as: (i) a service provider interested in expanding its service area without having to re-train a control policy from scratch, and (ii) when faced with extremely large urban networks, a service provider might consider training a policy on specific sub-graphs of the network, and later being able to deploy the learned policy on the entire system.

\subsection{Irregular Geographies}
\label{subsec:irregular_geographies}

\begin{table}[t]
\caption{System Performance on the Irregular 16-dimensional Topology (New York)}
\centering
\begin{tabular}{p{2cm} c c c}
    \hline
     & Reward & Served & Rebalancing  \\
     & (\%Dev. MPC-tri-level) & Demand & Cost (\$) \\ [0.25ex] 
    \hline
    ED & 7,900 (-27.9\%) & 2,431 & 3,451 \\ [0.2ex]
    A2C-GNN & 9,981 (\textbf{-8.9\%}) & 2,531 & 1,371 \\ [0.2ex]
    MPC-tri-level & 10,955 (0\%) & 2,527 & 382 \\ [0.2ex]
    MPC-standard & 10,127 (-7.6\%) & 2,376 & 379 \\
    \hline
    \end{tabular}%
  \label{tab:nyc_irregular}%
\end{table}

We now investigate how well the pre-trained A2C-GNN can be applied to arbitrary, non-grid-like transportation networks. 
Specifically, we select 16 stations defining a \emph{disjoint} service area, thus not representable as a contiguous grid over the city's geography.
Results in Table \ref{tab:nyc_irregular} show that the proposed approach achieves almost $40\%$ reduction in rebalancing cost together with a $26\%$ increase in profit compared to an equally distributed policy, thus exhibiting natural adaptation capabilities to irregular geographies.
Notice that MPC-standard performs worse than MPC tri-level in this case, as MPC approaches optimize system performance only within a time horizon, and if the time horizon is not sufficiently long, the resulting decisions can be suboptimal. 

Most importantly, the permutation of areas and their disjoint positioning, would make the A2C-MLP and A2C-CNN ill-defined for the task.
Namely, by being constrained to vector and grid representations of the urban topology, MLP-based and CNN-based agents could exhibit structural limitations when dealing with potentially complex networks not easily expressed in fixed representations. 
On the other hand, by explicitly representing stations as relational entities in a graph, graph neural networks enable reinforcement learning agents to recover extremely flexible behavior policies when dealing with diverse and irregular urban topologies.

\subsection{Network Granularity}
\label{subsec:network_granularity}

\begin{figure}[t]
      \centering
      \includegraphics[width=1\linewidth]{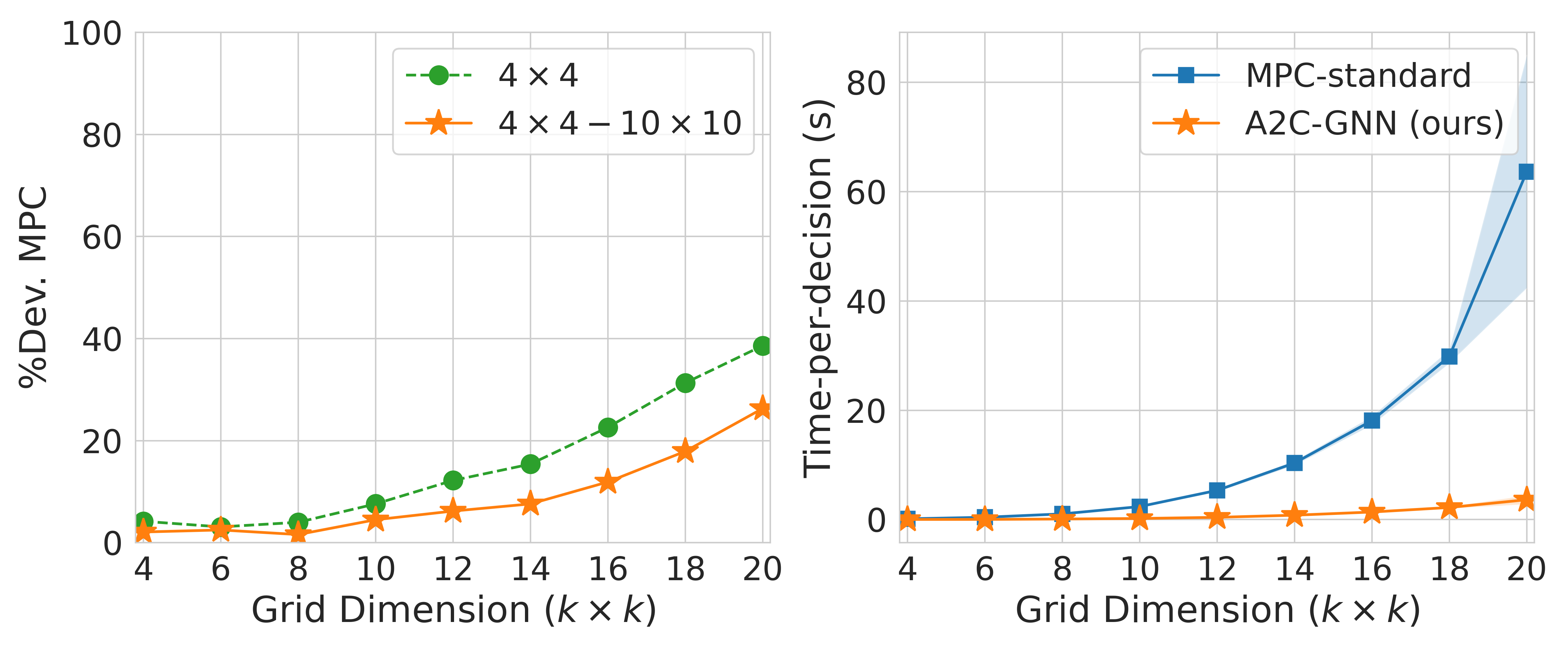}
      \caption{Left: System performance (Percentage Deviation from MPC-standard) for agents trained either on a single granularity ($4\times4$) or across granularities ($4\times4$ - $10\times10$), Right: Comparison of computation times between A2C-GNN and MPC-standard.}
      \label{fig:nyc_granularity}
  \end{figure}

From a service provider perspective, given a spatial segmentation of the service area, a critical decision is the one characterizing the spatial area associated with each node, or \emph{granularity}, of each rebalancing area.
To assess how well A2C-GNN is able to generalize across different service granularities we also study adaptation to finer spatial segmentations.
Specifically, we define the task as follows: given arbitrary granularities ranging from $4 \times 4$ to $20 \times 20$ grids (with $2 \times 2$ increments), we are interested in evaluating the portability of A2C-GNN when trained on coarse granularities and later applied, without any fine-tuning, to finer spatial scales.
Results in Fig. \ref{fig:nyc_granularity} (left) show transfer performance for A2C-GNN under two different training strategies.
The first pre-trains a rebalancing policy solely on a $4 \times 4$ grid, while the second exposes the agent to a diversity of granularities also at training time by considering grids up until $10 \times 10$.
The results show that training across granularities strongly increases the generalization capabilities of A2C-GNN, leading to rebalancing policies effective also under extreme granularity variations.
Crucially, we believe these results show clear evidence that it is possible to explicitly consider transfer and generalization in the design of the networks and training pipelines, and we believe this is an interesting and fruitful direction for future work.

\subsection{Computational analysis}
\label{subsec:computational_analysis}
Lastly, we study the computational cost of A2C-GNN compared to MPC-based solutions.
As shown in Fig. \ref{fig:nyc_granularity} (right), we compare the time necessary for both approaches to compute a single rebalancing decision.
Specifically, we do so across varying dimensions of the underlying transportation network, ranging from $16$ up until $400$ stations.
The results show that, once trained, learning-based approaches allow for fast computation of rebalancing policies by forward-propagation of the current system state through the learned policy $\pi_{\theta}(\ba_t | \bs_t)$.
Most importantly, A2C-GNN exhibits computational complexity linear in the number of nodes and graph connectivity, as opposed to control-based approaches which scale super-linearly in the number of edges \cite{Brand2019}.

\section{CONCLUSION}
\label{sec:conclusion}
This work addresses the problem of recovering effective rebalancing strategies for AMoD systems by proposing graph neural networks as a general approach to parametrize policy-based reinforcement learning agents.
We introduce an actor-critic algorithm where both policy and value function estimator make use of graph convolutions to define an agent capable of dealing with structured transportation networks. 
Our experiments focus on real-world case studies and show how the proposed architecture is able to achieve close-to-optimal performance on a variety of scenarios.
Crucially, we show how the relational inductive biases introduced by graph neural networks allow reinforcement learning agents to recover highly flexible, generalizable and scalable behavior policies.
In future work, we plan to investigate ways to explicitly consider transfer and generalization in the design of neural architectures and training strategies, such as casting the problem under the lenses of a Meta-Reinforcement Learning problem.
Another promising avenue of research is characterized by increasing the complexity and stochasticity in the system dynamics, such as considering a mixed fleet of autonomous and human driven vehicles.
Given their ability to learn about the system dynamics through interaction with an environment, we believe reinforcement learning approaches to be well suited for challenging, stochastic environments as the ones described by complex human-robot interactions.

\medskip

\bibliographystyle{unsrt} 
\bibliography{main,ASL_papers}

\newcommand{\noopsort}[1]{} \newcommand{\printfirst}[2]{#1}
  \newcommand{\singleletter}[1]{#1} \newcommand{\switchargs}[2]{#2#1}
\begin{thebibliography}{10}

\bibitem{UN2014}
{United Nations}.
\newblock World urbanization prospects: The 2014 revision.
\newblock Technical report, {United Nations}, 2014.

\bibitem{Pavone2015}
M.~Pavone.
\newblock {Autonomous} {Mobility-on-Demand} systems for future urban mobility.
\newblock In {\em Autonomes Fahren}. {Springer}, 2015.

\bibitem{ZhangPavone2016}
R.~Zhang and M.~Pavone.
\newblock Control of robotic {Mobility-on-Demand} systems: A
  queueing-theoretical perspective.
\newblock {\em {Int.\ Journal of Robotics Research}}, 35(1--3):186--203, 2016.

\bibitem{HylandMahmassani2018}
M.~Hyland and H.-S. Mahmassani.
\newblock Dynamic autonomous vehicle fleet operations: Optimization-based
  strategies to assign {AVs} to immediate traveler demand requests.
\newblock {\em {Transportation Research Part C: Emerging Technologies}},
  92:278--297, 2018.

\bibitem{LevinKockelmanEtAl2017}
M.~W. Levin, K.~M. Kockelman, S.~D. Boyles, and T.~Li.
\newblock A general framework for modeling shared autonomous vehicles with
  dynamic network-loading and dynamic ride-sharing application.
\newblock {\em Computers, Environment and Urban Systems}, 64:373 -- 383, 2017.

\bibitem{IglesiasRossiEtAl2018}
R.~Iglesias, F.~Rossi, K.~Wang, D.~Hallac, J.~Leskovec, and M.~Pavone.
\newblock Data-driven model predictive control of autonomous mobility-on-demand
  systems.
\newblock In {\em {Proc.\ IEEE Conf.\ on Robotics and Automation}}, 2018.

\bibitem{LeiQianEtAl2020}
Z.~Lei, X.~Qian, and S.-V. Ukkusuri.
\newblock Efficient proactive vehicle relocation for on-demand mobility service
  with recurrent neural networks.
\newblock {\em {Transportation Research Part C: Emerging Technologies}}, 117,
  2020.

\bibitem{GueriauCugurulloEtAl2020}
M.~Gu{\'e}riau, F.~Cugurullo, R.~Acheampong, and I.~Dusparic.
\newblock Shared {Autonomous Mobility on Demand}: A learning-based approach and
  its performance in the presence of traffic congestion.
\newblock {\em {IEEE Intelligent Transportation Systems Magazine}},
  12(4):208--218, 2020.

\bibitem{HollerVuorioEtAl2019}
J.~Holler, R.~Vuorio, Z.~Qin, X.~Tang, Y.~Jiao, T.~Jin, S.~Singh, C.~Wang, and
  J.~Ye.
\newblock Deep reinforcement learning for multi-driver vehicle dispatching and
  repositioning problem.
\newblock In {\em {IEEE Int.\ Conf.\ on Data Mining}}, 2019.

\bibitem{FluriRuchEtAl2019}
C.~Fluri, C.~Ruch, J.~Zilly, J.~Hakenberg, and E.~Frazzoli.
\newblock Learning to operate a fleet of cars.
\newblock In {\em {Proc.\ IEEE Int.\ Conf.\ on Intelligent Transportation
  Systems}}, 2019.

\bibitem{MnihKavukcuogluEtAl2015}
V.~Mnih, K.~Kavukcuoglu, D.~Silver, A.~Rusu, et~al.
\newblock Human-level control through deep reinforcement learning.
\newblock {\em {Nature}}, 518(7540):529--533, 2015.

\bibitem{SchulmanWolskiEtAl2017}
J.~Schulman, F.~Wolski, P.~Dhariwal, A.~Radford, and O.~Klimov.
\newblock Proximal policy optimization algorithms, 2017.
\newblock {Available at }\url{https://arxiv.org/abs/1707.06347}.

\bibitem{Williams1992}
R.-J. Williams.
\newblock Simple statistical gradient-following algorithms for connectionist
  reinforcement learning.
\newblock {\em {Machine Learning}}, 1992.

\bibitem{SuttonBarto1998}
R.~S. Sutton and A.~G. Barto.
\newblock {\em Reinforcement Learning: An Introduction}.
\newblock {MIT Press}, 1 edition, 1998.

\bibitem{LeCunBoserEtAl1989}
Y.~LeCun, B.~Boser, J.-S. Denker, D.~Henderson, R.-E. Howard, W.~Hubbard, and
  L.-D. Jackel.
\newblock Backpropagation applied to handwritten zip code recognition.
\newblock {\em {Neural Computation}}, 1(4):541--551, 1989.

\bibitem{BronsteinBrunaEtAl2017}
M.-M. Bronstein, J.~Bruna, Y.~LeCun, A.~Szlam, and P.~Vandergheynst.
\newblock Geometric deep learning: Going beyond euclidean data.
\newblock {\em {IEEE Signal Processing Magazine}}, 34(4):18–42, 2017.

\bibitem{KipfWelling2017}
T.-N. Kipf and M.~Welling.
\newblock Semi-supervised classification with graph convolutional networks.
\newblock In {\em {Int.\ Conf.\ on Learning Representations}}, 2017.

\bibitem{Daganzo1978}
C.-F Daganzo.
\newblock An approximate analytic model of many-to-many demand responsive
  transportation systems.
\newblock {\em {Transportation Research}}, 12(5):325--333, 1978.

\bibitem{GammelliEtAl2020}
D.~Gammelli, I.~Peled, F.~Rodrigues, D.~Pacino, and F.C. Pereira.
\newblock Estimating latent demand of shared mobility through censored gaussian
  processes.
\newblock {\em {Transportation Research Part C: Emerging Technologies}}, 120,
  2020.

\bibitem{MnihPuigdomenechEtAl2016}
V.~Mnih, A.~Puigdomenech, M.~Mirza, A.~Graves, T.-P. Lillicrap, T.~Harley,
  D.~Silver, and K.~Kavukcuoglu.
\newblock Asynchronous methods for deep reinforcement learning.
\newblock In {\em {Int.\ Conf.\ on Learning Representations}}, 2016.

\bibitem{ZhangRossiEtAl2016b}
R.~Zhang, F.~Rossi, and M.~Pavone.
\newblock Model predictive control of {Autonomous} {Mobility-on-Demand}
  systems.
\newblock In {\em {Proc.\ IEEE Conf.\ on Robotics and Automation}}, 2016.

\bibitem{TaxiLimousineCommission2013}
{Taxi \& Limousine Commission}.
\newblock {New York City Taxi \& Limousine Commission Trip Record Data}, 2013.
\newblock {See}
  \url{https://www1.nyc.gov/site/tlc/about/tlc-trip-record-data.page}.

\bibitem{Didi2016}
{ Didi Chuxing Technology Co.}
\newblock {Didi Chuxing Gaia Initiative}, 2016.
\newblock {Available at
  }\url{https://outreach.didichuxing.com/research/opendata/en/}.

\bibitem{PaszkeGrossEtAl2019}
A.~Paszke, S.~Gross, F.~Massa, A.~Lerer, et~al.
\newblock Pytorch: An imperative style, high-performance deep learning library,
  2019.
\newblock {Available at }\url{https://arxiv.org/abs/1912.01703}.

\bibitem{ios_ILOG:1987}
IBM.
\newblock {\em {ILOG CPLEX} User's guide}.
\newblock IBM ILOG, 1987.

\bibitem{KingmaBa2015}
D.~Kingma and J.~Ba.
\newblock Adam: A method for stochastic optimization.
\newblock In {\em {Int.\ Conf.\ on Learning Representations}}, 2015.

\bibitem{Brand2019}
J.~van~den Brand.
\newblock A deterministic linear program solver in current matrix
  multiplication time.
\newblock In {\em {ACM-SIAM Symp.\ on Discrete Algorithms}}, 2020.

\end{thebibliography}

\addtolength{\textheight}{-12cm}  

\end{document}